\documentclass[10pt,aps,prd,reprint,nofootinbib,floats,floatfix,amsfonts,amssymb,amsmath,preprintnumbers,notitlepage,superscriptaddress]{revtex4-1}
\usepackage{aastex_macros}
\usepackage{float}
\usepackage{graphicx}
\usepackage{caption}
\usepackage{subcaption}
\usepackage[utf8]{inputenc}
\usepackage[british]{babel}
\usepackage{isomath}
\usepackage[protrusion=true,tracking=true,kerning=true,spacing=true,final,babel=true]{microtype}
\usepackage{physics}
\usepackage{amsmath}
\usepackage{xcolor}
\usepackage{url}
\usepackage[final]{hyperref}

\usepackage{multirow}

\begin{document}

%\begin{flushleft}
%KCL-PH-TH-2021-93 
%\vspace{-0.3cm}
%\end{flushleft}

\title{Detecting parity violation from axion inflation with third generation detectors}
\author{Charles Badger}
\email{charles.badger@kcl.ac.uk}
\affiliation{Theoretical Particle Physics and Cosmology Group, \, Physics \, Department, \\ King's College London, \, University \, of London, \, Strand, \, London \, WC2R \, 2LS, \, UK}
\author{Mairi Sakellariadou}
\email{mairi.sakellariadou@kcl.ac.uk}
\affiliation
{Theoretical Particle Physics and Cosmology Group, \, Physics \, Department, \\ King's College London, \, University \, of London, \, Strand, \, London \, WC2R \, 2LS, \, UK}

\begin{abstract}
A gravitational wave background is expected to emerge from the superposition of numerous gravitational wave sources of both astrophysical and cosmological origin. A number of cosmological models can have a parity violation, resulting in the generation of circularly polarised gravitational waves. We investigate the constraining power of third generation Einstein Telescope and Cosmic Explorer detectors, for a gravitational wave background generated by early universe axion inflation. 
\end{abstract}

\maketitle

\emph{Introduction---}
A stochastic gravitational wave background (SGWB) is expected to be created from the overlap of gravitational waves (GWs) coming from many independent sources. A number of early universe cosmological sources leading to a SGWB have been proposed, including inflation~\cite{AxInf_GW}, cosmic strings~\cite{CosStr_GW}, first order phase  transitions (see, e.g.~\cite{Caprini:2015zlo,Hindmarsh:2020hop}), or cosmological models inspired from string theory (see, e.g.~\cite{altCos_GW,altCos2_GW}). 

A number of mechanisms in the early universe can create parity violation \cite{Alexander_2006} that may manifest itself in the production of asymmetric amounts of right- and left-handed circularly polarised isotropic GWs. A detection and subsequent analysis of a polarised SGWB can place constraints on parity violating theories. Searches for parity violation in the LIGO-Virgo data has been  explored \cite{Crowder_2013, PV_Turb}.

In this study we focus on parity violation effects from axion inflation sourced GWs  (e.g.,~\cite{AxInf_PV, AxInf_PV_Gauss, GW_AxInf}) in the context of the upcoming 3rd generation (3g) detectors Einstein Telescope (ET)~\cite{ET_Website} and Cosmic Explorer (CE)~\cite{reitze2019cosmic}.
We adopt the formalism~\cite{PV_Turb} presented in~\cite{Formalism_MathMod}. We first highlight the methodology and then apply it to a parity violating axion inflation model.\\

\emph{Method---}
\label{sec: methods}
We perform parameter estimation and fit GW models to data using a hybrid frequentist-Bayesian approach~\cite{Matas:2020roi}. We construct a Gaussian log-likelihood for a multi-baseline network
\begin{align}
    \log p(\hat C(f) | \boldsymbol{\theta})
    \propto \sum_{d_1 d_2}\sum_{f}\frac{\left[\hat C_{d_1 d_2}(f) - \Omega'_{\rm GW}(f, \boldsymbol{\theta}) \right]^2}{\sigma_{d_1 d_2}^2(f)}~;
    \label{eq:pe_ms:likelihood}
\end{align}
$\hat{C}_{d_1 d_2}(f)$
is the frequency-dependent cross-correlation estimator of the SGWB  for detectors $d_1, d_2$, and $\sigma^2_{d_1 d_2}(f)$  its variance~\cite{Formalism_Source} for model parameters $\boldsymbol{\theta}$. The cross-correlation statistics are constructed using strain data from the individual  detectors. We assume that correlated noise sources have been either filtered out~\cite{Cirone:2018guh} or accounted for~\cite{Meyers:2020qrb}. 
The normalised GW energy density model we fit to the data is 
\begin{align}
    \Omega'_{\rm GW}(f, \boldsymbol{\theta}) = \Omega_{\rm GW}(f, \boldsymbol{\theta})\bigg[1+\Pi(f)\frac{\gamma_V^{d_1 d_2}(f)}{\gamma_I^{d_1 d_2}(f)}\bigg]~;
    \label{OmegaGW_PiForm}
\end{align}
$\gamma_I^{d_1 d_2}$ stands for the standard overlap reduction function of  detectors $d_1, d_2$, and $\gamma_V^{d_1 d_2}$ denote the overlap function associated with the parity violation term~\cite{Romano_Cornish}. The polarisation degree, $\Pi(f)$, takes on values between -1 (fully left polarisation) and 1 (fully right polarisation), with $\Pi = 0$ for unpolarised isotropic SGWB. [More details can be seen in Appendix \ref{app: Methods}.]\\

\emph{Axion Inflation---}
\label{sec: models}
Consider a pseudoscalar inflaton field $\phi$ coupled to $\mathcal{N}$ U(1) gauge fields $A^a_{\mu}$ as \cite{PhysRevD.37.2743, Prim_Goldstone, Anber_2006}
\begin{align}
    \mathcal{L} = -\frac{1}{2}\partial_{\mu}\phi \partial^{\mu}\phi - \frac{1}{4}F^a_{\mu \nu}F_a^{\mu \nu} - V(\phi) - \frac{\alpha^a}{4\Lambda}\phi F^a_{\mu \nu}\Tilde{F_a^{\mu \nu}}~;
    \label{eq: L_background}
\end{align}
$F^a_{\mu \nu}$ ($\Tilde{F_a^{\mu \nu}}$) is the (dual) field strength tensor, $\Lambda$ is the mass scale suppressing higher dimensional operators of the theory, and $\alpha$ parametrises the strength of the inflaton gauge field coupling, $\alpha^a = \alpha$ for all $a\leq \mathcal{N}$. 
The resulting equations of motion imply
\begin{align}
     -\phi_{,N} + M_{\rm Pl}^2\frac{V_{,\phi}}{V} = \mathcal{N}\times\frac{2.4}{9M_{\rm Pl}^2} \times 10^{-4} \Big(\frac{\alpha}{\Lambda}\Big)\frac{V}{\xi^4} e^{2\pi\xi}~;
     \label{eq: numericalCalc}
 \end{align}
where $\xi$ is defined as
\begin{align}
    \xi \equiv \frac{\alpha |\dot{\phi}|}{2\Lambda H}~;
    \label{eq: xi_def}
\end{align}
$V(\phi)$ stands for the potential of the inflaton field, $N$ is the number of e-folds, we have partial derivatives $\phi_{,N} = \partial\phi / \partial N$ and $V_{,\phi} = \partial V / \partial \phi$, and dot denotes a derivative with respect to cosmic time $t$. Equation~(\ref{eq: numericalCalc}), obtained under the assumption  $\phi > 0$, $V_{,\phi}>0$, $\dot{\phi}<0$, can be then used to numerically calculate the evolution of $\phi$ in terms of frequency $f$ as~\cite{AxInf_Nums}
\begin{align}
     N = N_{\rm CMB} + \ln{\frac{k_{\rm CMB}}{0.002\text{ Mpc}^{-1}}} - 44.9 - \ln{\frac{f}{10^2\text{ Hz}}}~;
     \label{eq: N_to_f}
 \end{align}
$k_{\rm CMB} = 0.002\text{ Mpc}^{-1}$ ~\cite{AxInf_Nums} and $N_{\rm CMB}\approx (50-60)$ defined as the total number of e-folds after the CMB scales exited the horizon. 
In terms of the number of e-folds, $\xi$ can be written as $\xi= \alpha /( 2\Lambda)\phi_{,N}$.

Given a scalar potential $V(\phi)$, one can use Planck 2018 data~\cite{Planck_2018} to impose parameter constraints.
A well-approximated solution for tensor and scalar perturbations reads~\cite{AxInf_PV_Gauss, PBH_AxInf_Upper}
\begin{align}
    \Omega_{\rm GW} \simeq \frac{1}{12}\Omega_{R,0}\Big(\frac{V(\phi)}{3\pi^2 M_{\rm Pl}^4}\Big) \Big(1 + 4.3\times 10^{-7}\mathcal{N}\frac{V(\phi)}{3M_{\rm Pl}^4 \xi^6} e^{4\pi\xi} \Big)~,
    \label{eq: GW_Eq}
\end{align}
and
%\begin{multline}
%    \Delta_{\rm s}^2 
%    \simeq \Big(\frac{\sqrt{3V}}{12\pi\xi}M_{\rm Pl}\frac{\alpha}{\Lambda}\Big)^2 \\ 
%    + \Big(\frac{\alpha}{\Lambda}\Big)^4 \Big(\frac{M_{\rm Pl}}{2\beta\xi V\sqrt{\mathcal{N}}} \Big)^2 \Big(\mathcal{N}\times 2.4 \times 10^{-4} \frac{V^2}{9\xi^4} e^{2\pi\xi}\Big)^2~,
%    \label{eq: deltas_Eq}
%\end{multline}
\begin{multline}
    \Delta_{\rm s}^2 
    \simeq \frac{V(\phi)}{3}\bigg[\frac{1}{4\pi M_{\rm{Pl}}\xi}\Big(\frac{\alpha}{\Lambda}\Big)\bigg]^2 \\
    + \mathcal{N}\bigg[2.4\times 10^{-4} \Big(\frac{\alpha}{\Lambda}\Big)^2 \frac{V(\phi)}{3M_{\rm{Pl}}^2}\frac{e^{2\pi\xi}}{6\beta\xi^5}\bigg]^2,
    \label{eq: deltas_Eq}
\end{multline}
with %$$\beta \equiv 1 + 4.8 \times 10^{-4}\pi\big(\frac{\alpha}{\Lambda}\big)^2\mathcal{N} e^{2\pi\xi} \frac{M_{\rm Pl}V}{18\xi^4} ~,$$
$$\beta \equiv 1 + \mathcal{N} \frac{2.4 \times 10^{-4}}{9M_{\rm{Pl}}^2}\pi\Big(\frac{\alpha}{\Lambda}\Big)^2 \frac{V(\phi)}{\xi^4} e^{2\pi\xi} ~,$$
$\Omega_{R,0}=8.6\times 10^{-5}$ being the radiation energy density today and $M_{\rm Pl}$ the reduced Planck mass set to unity.
 
There is an additional constraint one however should take into account.
Density perturbations produced during inflation could collapse and form primordial black holes (PBHs), leading to a possible risk of PBH overproduction ~\cite{PBH_UpperData, PBH_UpperTools, PBH_AxInf_Upper}. The upper bound on scalar power spectrum $\Delta_{\rm s}^2$ from the non-detection of PBHs is $\Delta_s^2 \approx 10^{-4}$~\cite{PBH_Workarounds}.

\begin{figure}
    \centering
    \includegraphics[width=0.5\textwidth]{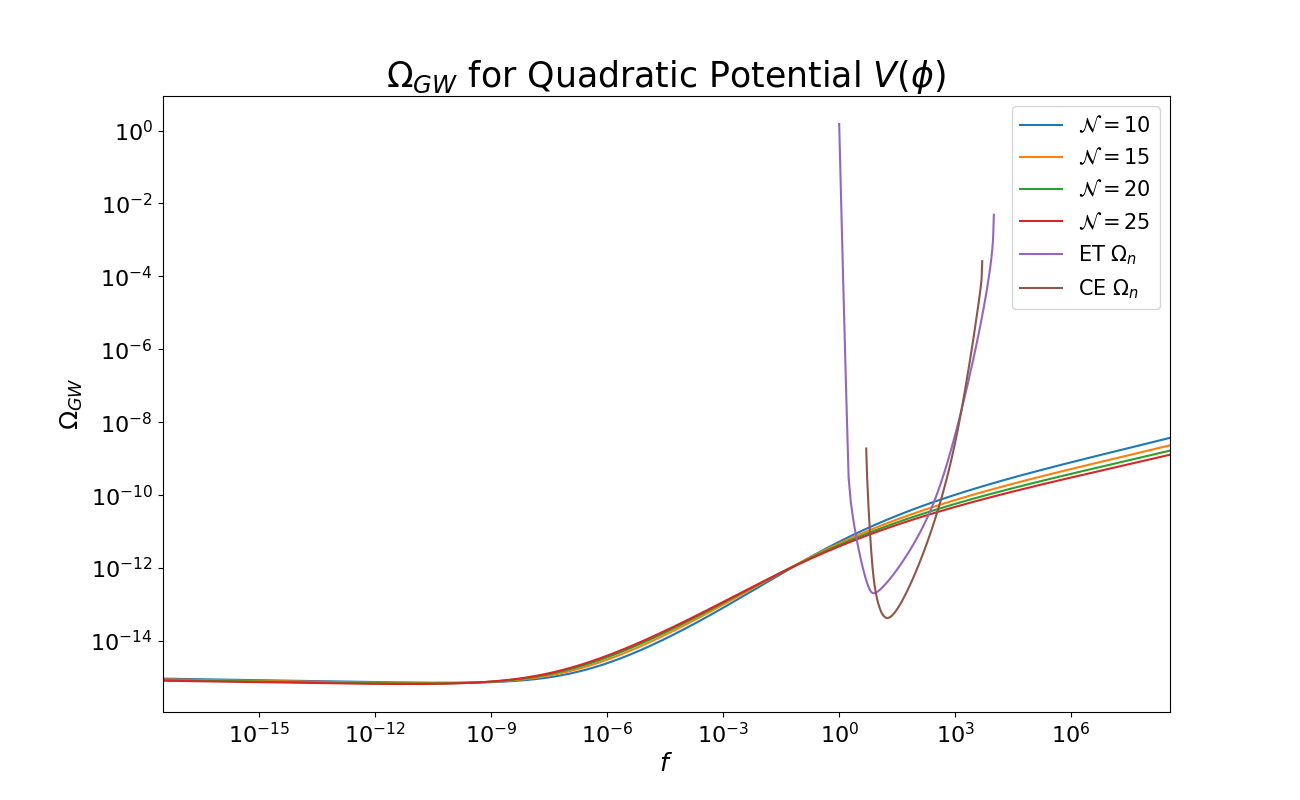}
    \caption{$\Omega_{\rm GW}(f)$ of quadratic model with $\xi_{\rm CMB} = 2.5$, $\mathcal{N} = 10, 15, 20, 25$ plotted with Einstein Telescope and Cosmic Explorer noise density $\Omega_n$.}
    \label{fig: OmegaGW_QuadPlot}
\end{figure}

 Among the different axion inflation models, let us consider the well-motivated quadratic model \cite{LINDE1983177, 10.1143/ptp/86.1.103, GravLin, Nflation}
 \begin{align}
     V(\phi) = \lambda\phi^2~,
     \label{eq: Vmodel_power}
 \end{align}
 from the chaotic potential $V \sim \phi^n$ class~\cite{LINDE1983177}. 
 Resulting $\Omega_{\rm GW}$ spectra with maximum strength $\xi_{\rm CMB} \equiv \xi(N_{\rm CMB}) = 2.5$ ($\alpha / \Lambda \simeq 39$) \cite{AxInf_Nums} and gauge fields $\mathcal{N} = 10, 15, 20, 25$ are shown in Fig.~\ref{fig: OmegaGW_QuadPlot}.

\begin{figure}
    \centering
    \includegraphics[width=0.5\textwidth]{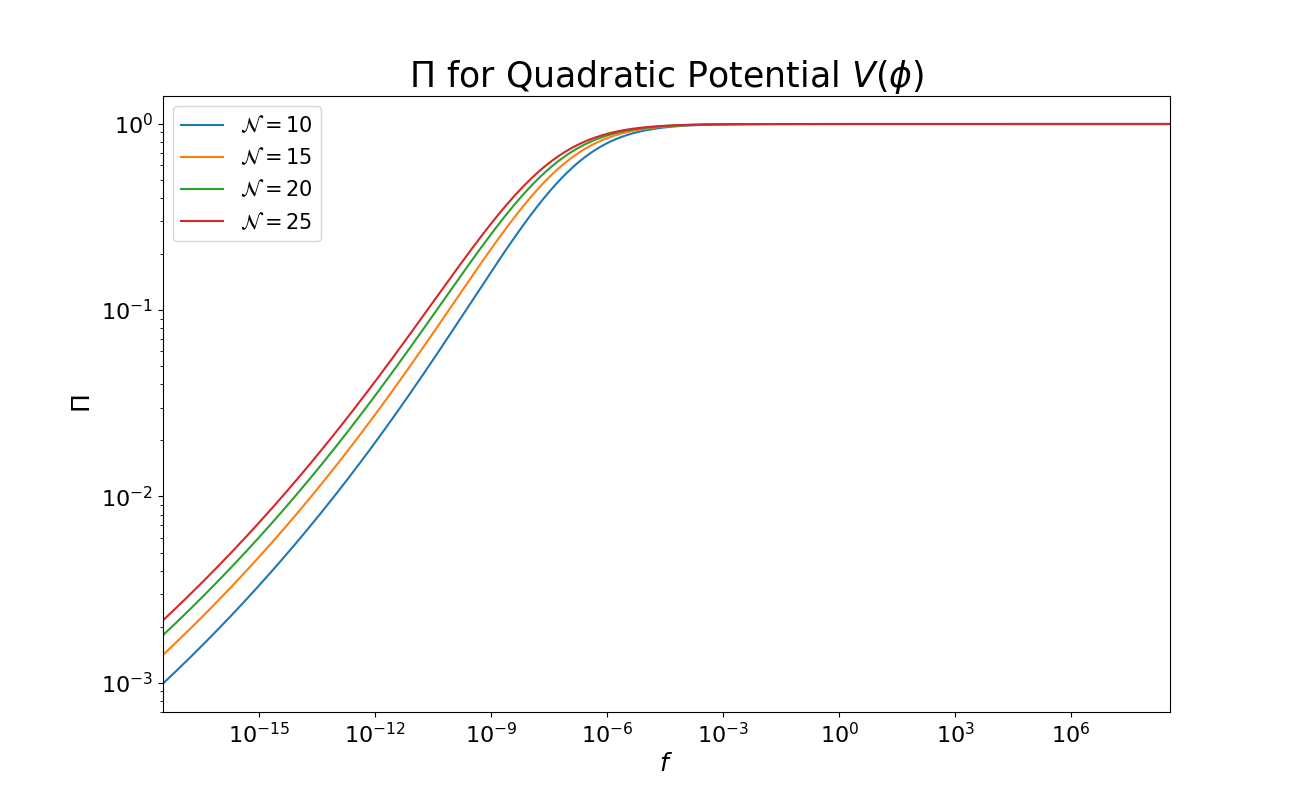}
    \caption{$\Pi(f)$ of quadratic model with $\xi_{\rm CMB} = 2.5$, $\mathcal{N} = 10, 15, 20, 25$.}
    \label{fig: Pi_QuadPlot}
\end{figure}

The corresponding polarisation degree reads~\cite{AxInf_PV}:
\begin{align}
    \Pi \simeq \frac{4.3\times 10^{-7}\frac{\lambda\phi^2}{3M_{\rm Pl}^4}\frac{e^{4\pi\xi}}{\xi^6}}{1 + 4.3\times 10^{-7}\frac{\lambda\phi^2}{3M_{\rm Pl}^4}\frac{e^{4\pi\xi}}{\xi^6}}.
    \label{eq: PV_Pi}
\end{align}
We show in Fig.~\ref{fig: Pi_QuadPlot} the polarisation degree $\Pi$ for GW spectra with $\xi_{\rm CMB} = 2.5$ and $\mathcal{N} = 10, 15, 20, 25$.

%Since we focus on 3g detectors ($5$ Hz $\leq f \leq 5000$ Hz), almost entirely right-handed polarisation (approximately constant $\Pi \simeq 1$) is expected, as one can see from  Fig.~\ref{fig: Pi_QuadPlot}.

For $f \geq 10^{-3}$ Hz, almost entirely right-handed polarisation (approximately constant $\Pi \simeq 1$) is expected, as one can see from  Fig.~\ref{fig: Pi_QuadPlot}.

Analysing a LIGO and Virgo network with A+ noise sensitivity design~\cite{A+_Sens} we have found that such a configuration could not provide any promising results, hence we will consider a 3g network. \\

\begin{figure}
    \centering
    \includegraphics[width=0.45\textwidth]{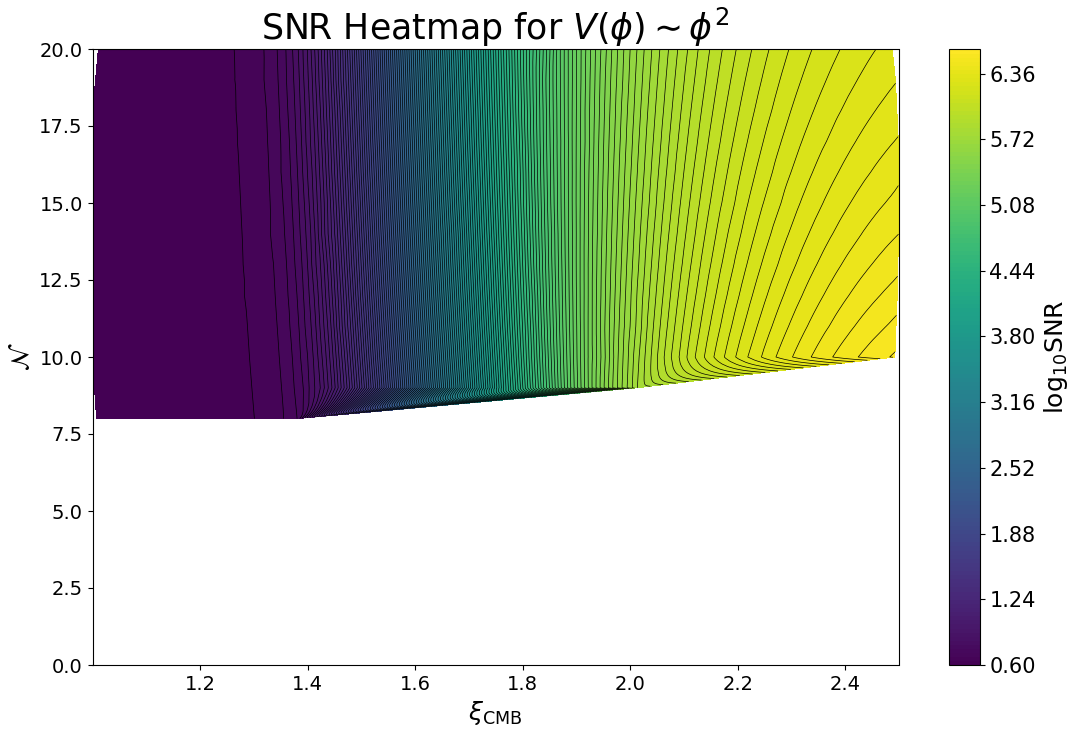}
    \caption{Heatmap $\rm{log}_{10}\rm{SNR}$ plotted for sampled $\mathcal{N}$ and $\xi_{\rm CMB}$ for assumed axion inflation quadratic model.}
    \label{fig: SNR_Heatmap}
\end{figure}

\emph{Results---}
\label{sec: results}
Using ET and CE noise sensitivity curves~\cite{ET_CE_Sens}, we construct a combined noise energy density $\Omega_n(f)$~\cite{Romano_Cornish}.
For the quadratic model we are focusing on, we consider 3500 random samples of discrete number of gauge fields $1 \leq \mathcal{N} \leq 25$, assuming also $1 \leq \xi_{\rm CMB} \leq 2.5$. For these samples we then calculate the corresponding spectra using Eqs.~(\ref{eq: GW_Eq}) and (\ref{eq: deltas_Eq}). If the resulting $\Delta_{\rm s}^2$ is below the PBH upper limit, we compute the corresponding signal-to-noise (SNR) ratio assuming $T = 3$ years of observation~\cite{Romano_Cornish}. Our results, shown  in Fig.~\ref{fig: SNR_Heatmap},
imply that quadratic models with 
%$\alpha / \Lambda \gtrsim 33$ 
$\xi_{\rm CMB} \gtrsim 2.0$ and $\mathcal{N} \geq 10$ yield strong GW spectra with $\rm{log}_{10}\rm{SNR} \gtrsim 5.5$ (all below the PBH upper limit).

\begin{figure}
    \centering
    \includegraphics[width=0.5\textwidth]{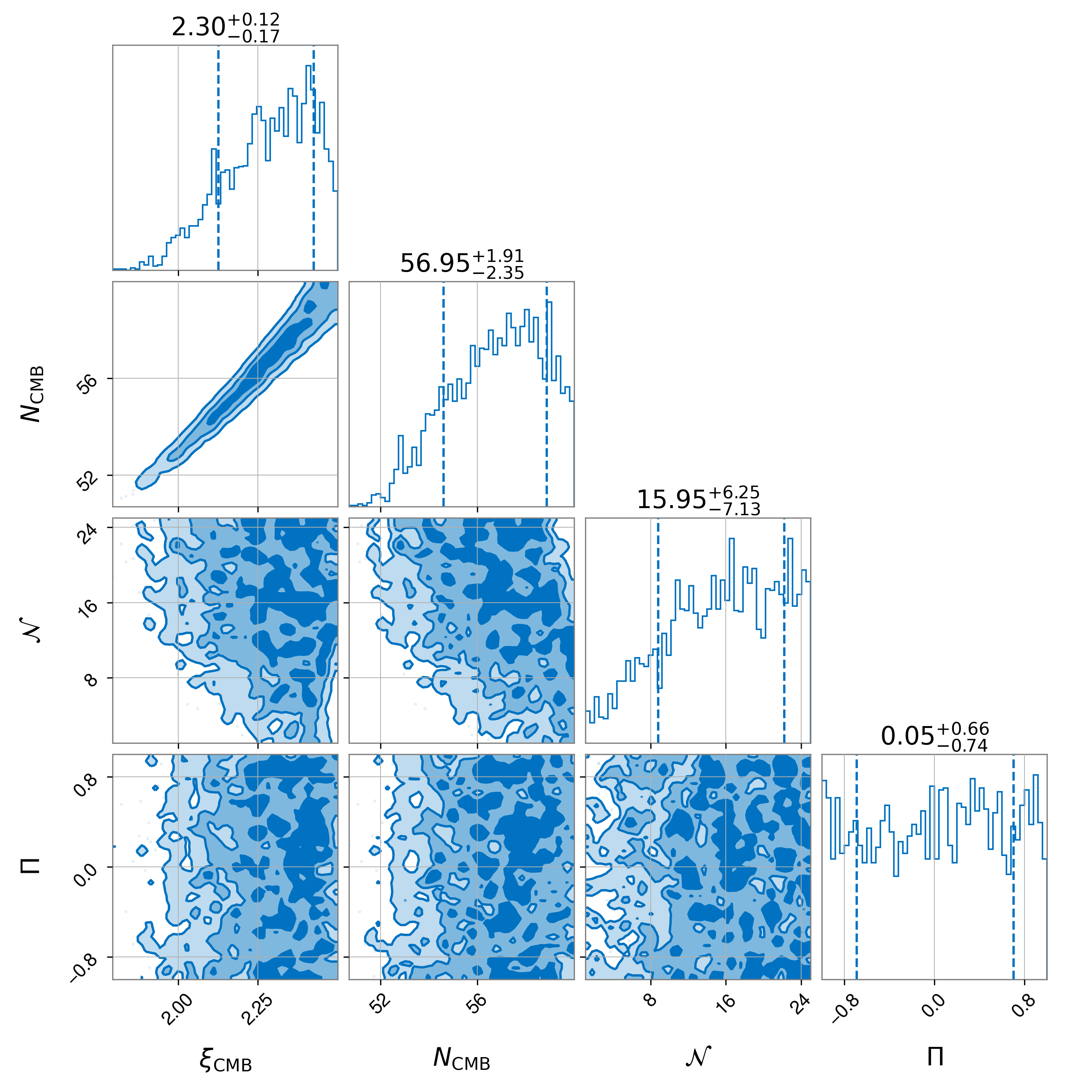}
    \caption{Parameter estimation corner plot of $\xi_{\rm{CMB}} = 2.5$ quadratic axion inflation using ET alone. Analysis found $\ln{\mathcal{B}_{\rm{Noise}}^{\rm{Quadratic}}} = 80.951 \pm 0.123$.}
    \label{fig: MaxET_corner}
\end{figure}

Let us first consider a triangular, three interferometer (60 degree opening angles, each separated by 10 km) design for the ET network located at the Virgo detector site. We inject a GW signal assuming an observation time of 3 years, $\xi_{\rm CMB}=2.5$, and $\mathcal{N}=10$.
We search uniformly for $1 \leq \xi_{\rm CMB} \leq 2.5$, number of U(1) gauge fields $1 \leq \mathcal{N} \leq 25$, total e-folds $50 \leq N_{\rm CMB} \leq 60$ 
 and parity violation parameter $-1 \leq \Pi \leq 1$. 
 %The corner plot is shown in Fig.~\ref{fig: MaxET_corner}. 
 The corner plot of the posterior distribution is shown in Fig.~\ref{fig: MaxET_corner}.
 It is clear that while reasonable constraints can be placed on %$\alpha / \Lambda, N_{\rm CMB} \text{ and } n$
$\xi_{\rm CMB}$ and $N_{\rm CMB}$, this is not the case for $\mathcal{N} \text{ and } \Pi$. Poor constraints on $\mathcal{N}$ are expected since different values of $\mathcal{N}$ lead to similar $\Omega_{\rm GW}(f)$ spectra within the relevant ET frequency range (see Fig.~\ref{fig: OmegaGW_QuadPlot}). An inability to place constraints on $\Pi$ is also clear since for the ET network, two ET interferometers have $\gamma_{V}^{ET_i ET_j}(f) \approx 0$ resulting in $\Omega_{\rm GW}^{'} \simeq \Omega_{\rm GW}$ for any value of $\Pi$ when using this formalism. 
The Bayes factor  $\ln{\mathcal{B}_{\rm{Noise}}^{\rm{Quadratic}}} = 80.951 \pm 0.123$ clearly indicates a preference for the quadratic model with respect to noise.

To improve the parameter estimation, let us add additional 3g detectors. As one can see from Fig.~\ref{fig: SNR_Heatmap}, a large SNR could be obtained with $\xi_{\rm{CMB}} \gtrsim 2.0$ and $\mathcal{N} \geq 10$.
We thus perform 500 Monte Carlo GW injections with $2.0 \leq \xi_{\rm CMB} \leq 2.5$, 
%{\bf (justify the different choice for the range of $\xi$)}.  
assuming $\mathcal{N}=10$ and $N_{\rm CMB} = 60$. The choice of $\mathcal{N}=10$ is made such that we have the strongest signal (see, Fig.~\ref{fig: OmegaGW_QuadPlot}).
We analyse our results using one ET located at the Virgo site, with either one Cosmic Explorer (CE) located at the LIGO Hanford detector site~\cite{TheLIGOScientific:2014jea}, or two CEs located at the LIGO Hanford and Livingston sites, respectively.

\begin{figure}
    \centering
    \includegraphics[width=0.5\textwidth]{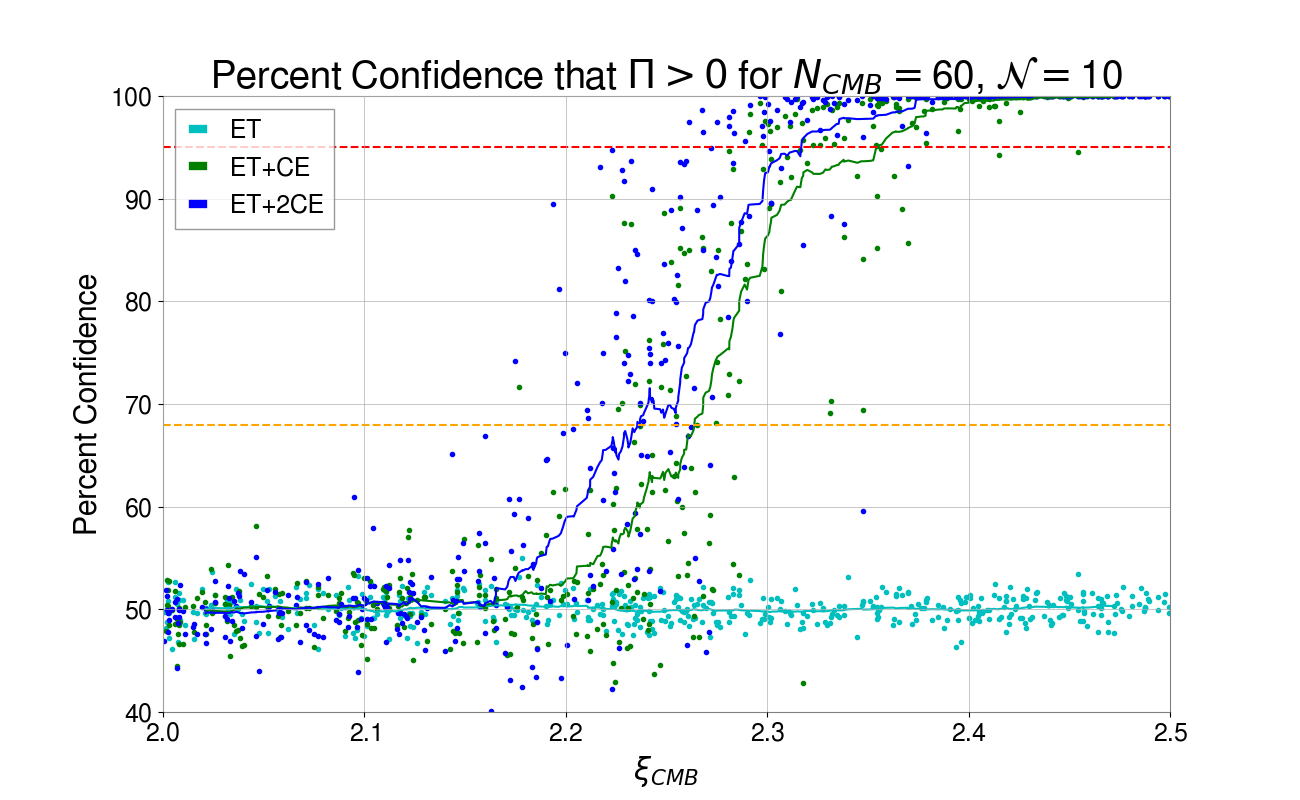}
    \caption{Percent Confidence that $\Pi > 0$ for detector networks ET, ET + CE and ET + 2 CEs.}
    \label{fig: Pi_Conf}
\end{figure}

In Fig.~\ref{fig: Pi_Conf} we plot the percent confidence that $\Pi > 0$ based off of each injection's results \footnote{Percent confidence that $\Pi > 0$ is the proportion of the $\Pi$ posterior distribution that is greater than 0.}. We list in Table \ref{table: Pi_Conf} the injected $\xi_{\rm{CMB}}$ for which 68\%, 95\% and 99.7\% confidence can be achieved. %thresholds that $\Pi > 0$.

\begin{table}[]
\begin{tabular}{|p{3cm}|p{2cm}|p{2cm}|}
\hline
\multicolumn{1}{|c|}{\textbf{Confidence $(\Pi > 0)$}} & \multicolumn{1}{c|}{\textbf{ET + CE}} & \multicolumn{1}{c|}{\textbf{ET + 2 CEs}} \\ \hline
\begin{tabular}[c]{@{}l@{}} 68\%\end{tabular} & 2.26 & 2.24 \\ \hline
95\% & 2.35 & 2.32 \\ \hline
99.7\% & 2.43 & 2.39 \\ \hline
\end{tabular}
\caption{$\xi_{\rm{CMB}}$ needed to claim $\Pi > 0$ at differing percent confidences for the respective detector networks.}
\label{table: Pi_Conf}
\end{table}

Figure~\ref{fig: Pi_Conf} clearly shows that additional 3g detectors improve our parity violation detection outlook. For quadratic models with $\xi_{\rm CMB} \gtrsim 2.35$, one can claim $\Pi > 0$ with at least 95\% certainty using just two 3g detectors.% are sufficient. 

\begin{figure}
    \centering
    \includegraphics[width=0.5\textwidth]{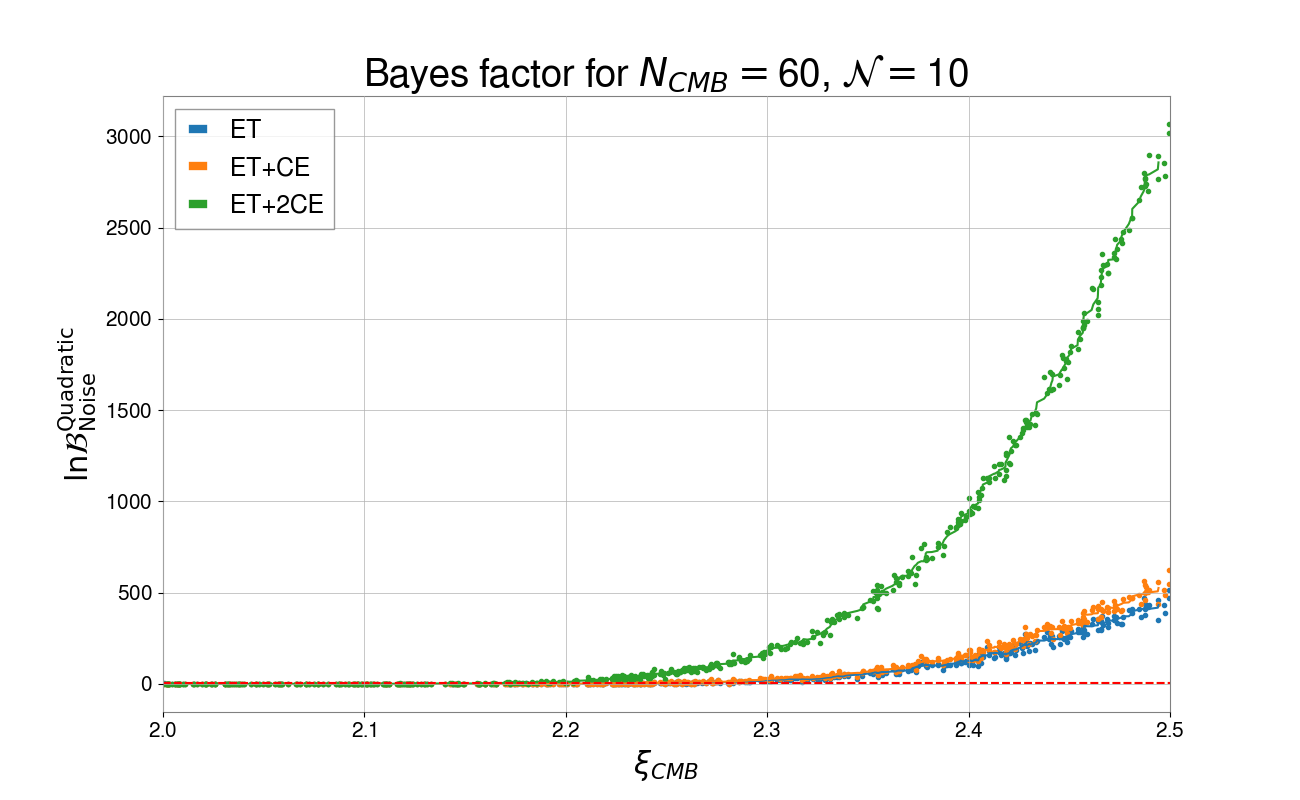}
    \caption{Bayes factor $\ln{\mathcal{B}_{\rm{Noise}}^{\rm{Quadratic}}}$ for ET, ET + CE and ET+ 2 CEs detector networks.}
    \label{fig: Bayes_Plot}
\end{figure}

We plot in Fig.~\ref{fig: Bayes_Plot} the Bayes factors $\ln{\mathcal{B}_{\rm{Noise}}^{\rm{Quadratic}}}$ for the 500 injections. 
It is clear that the strength of $\ln{\mathcal{B}_{\rm{Noise}}^{\rm{Quadratic}}}$ improves drastically with each additional CE detector in the network.
Strong preference ($\ln{\mathcal{B}_{\rm{Noise}}^{\rm{Quadratic}}} > 15$) for our parity violating quadratic potential model can be achieved with $\xi_{\rm CMB} \gtrsim 2.30$.

It is worth noting that our analysis assumed a total number of e-folds $N_{\rm{CMB}} = 60$. A smaller $N_{\rm{CMB}}$ would generate stronger $\Omega_{\rm{GW}}$ in the 3g detector frequency range. Thus, our results must be seen as a conservative insight for the quadratic potential model.\\

\emph{Conclusions---}
\label{sec: concl}
%\textbf{Beef up a little}
We have studied detection of parity violation with 3g detectors sourced from axion inflation focusing on the quadratic potential.

Using this model, we showed that we can avoid the overproduction of PBHs by considering at least 10 U(1) gauge field couplings. Large SNR $(\log_{10}\rm{SNR} \gtrsim 5.5)$ was obtained when $\xi_{\rm{CMB}} = \xi(N_{\rm{CMB}}) \gtrsim 2.0$. 

We showed that a SGWB with large $\xi_{\rm{CMB}} = 2.5$ examined by a ET network alone can constrain $\xi_{\rm{CMB}}$ and $N_{\rm{CMB}}$ reasonably well, but it is unable to constrain 
%the number of gauge fields $\mathcal{N}$ and 
the polarisation degree of the spectrum $\Pi$. Despite this, a Bayes factor  $\ln{\mathcal{B}_{\rm{Noise}}^{\rm{Quadratic}}} = 80.951 \pm 0.123$ was obtained in this analysis - indicating a strong preference for the quadratic axion inflation model with respect to noise.

Adding additional CE detectors to the network, we can better constrain the polarisation degree. With two 3g detectors (ET and CE) alone, one can claim with at least 95\% confidence that $\Pi > 0$ when $\xi_{\rm{CMB}} \gtrsim 2.35$. However, a network of three 3g detectors (ET and 2 CEs) is needed in order to make a confident claim about the detection of a quadratic axion inflation signature. Each additional CE detector added to the 3g network results in a drastic improvement in retrieved $\ln{\mathcal{B}_{\rm{Noise}}^{\rm{Quadratic}}}$.

\acknowledgments
We thank LIGO-Virgo-KAGRA collaboration stochastic members for discussions related to axion inflation.
%\textbf{Too generic?}
We acknowledge computational resources provided by the LIGO Laboratory and supported by National Science Foundation Grants PHY-0757058 and PHY-0823459. 
This paper has been given LIGO DCC number LIGO-P2100430, an Einstein Telescope (ET) documentation number ET-0457A-21, and a KCL TPPC number 2021-93.

M.S. is supported in part by the Science and Technology Facility Council (STFC), United Kingdom, under the research grant ST/P000258/1.

Software packages used in this paper are \texttt{matplotlib}~\cite{Hunter:2007}, \texttt{numpy}~\cite{numpy}, \texttt{bilby}~\cite{Ashton:2018jfp}, \texttt{ChainConsumer}~\cite{Hinton2016}.

\appendix
\section{Methods}
\label{app: Methods}
We normalize GW energy density for our data with
\begin{align}
    \Omega'_{\rm GW}(f, \boldsymbol{\theta}) = \Omega_{\rm GW}(f, \boldsymbol{\theta})\bigg[1+\Pi(f)\frac{\gamma_V^{d_1 d_2}(f)}{\gamma_I^{d_1 d_2}(f)}\bigg]~;
    \label{eq: PVOmega_App}
\end{align}
using model parameters $\boldsymbol{\theta}$, where we denote $\gamma_I^{d_1 d_2}$ as the standard overlap reduction function of two detectors $d_1, d_2$, and $\gamma_V^{d_1 d_2}$ as the overlap function associated with the parity violation term defined as:
\begin{eqnarray}
    \gamma_I^{d_1 d_2}(f) = \frac{5}{8\pi}\int d\hat{\Omega}(F_{d_1}^{+}F_{d_2}^{+*} + F_{d_1}^{\cross}F_{d_2}^{\cross*})e^{2\pi if\hat{\Omega}\cdot\Delta\Vec{x}}~, \nonumber\\
    \gamma_V^{d_1 d_2}(f) = -\frac{5}{8\pi}\int d\hat{\Omega}(F_{d_1}^{+}F_{d_2}^{\cross*} - F_{d_1}^{\cross}F_{d_2}^{+*})e^{2\pi if\hat{\Omega}\cdot\Delta\Vec{x}}~;
    \label{eq: ORF_App}
\end{eqnarray}
where $F_n^A = e_{ab}^A d_n^{ab}$ stands for the contraction of the tensor modes of polarisation $A$ to the $n^{\rm th}$ detector's geometry. The polarisation degree, $\Pi(f) = V(f)/I(f)$, takes on values between -1 (fully left polarisation) and 1 (fully right polarisation), with $\Pi = 0$ being an unpolarised isotropic SGWB.

To proceed we perform parameter estimation and fit GW models to data using a hybrid frequentist-Bayesian approach \cite{Matas:2020roi}. We construct a Gaussian log-likelihood for a multi-baseline network
\begin{align}
    \log p(\hat C(f) | \boldsymbol{\theta})
    \propto \sum_{d_1 d_2}\sum_{f}\frac{\left[\hat C_{d_1 d_2}(f) - \Omega'_{\rm GW}(f, \boldsymbol{\theta}) \right]^2}{\sigma_{d_1 d_2}^2(f)}~,
    \label{eq: likelihood_App}
\end{align}
where $\hat{C}_{d_1 d_2}(f)$ is the frequency-dependent cross-correlation estimator of the SGWB  for detectors $d_1, d_2$, and $\sigma^2_{d_1 d_2}(f)$ is its variance ~\cite{Formalism_Source}. We assume that correlated noise sources have been either filtered out \cite{Cirone:2018guh} or accounted for \cite{Meyers:2020qrb}. The normalised GW energy density model we fit to the data is $\Omega'_{\rm GW}(f,  {\boldsymbol \theta})$, with parameters $\boldsymbol\theta$ including both GW parameters as well as parameters of the $\Pi(f)$ model.

%\printbibliography
%\bibliographystyle{stylename}
%\bibliographystyle{unsrt}
%\bibliography{parity}

\end{document}